\documentclass[usenatbib, usepslatex]{mn2e}
\usepackage{txfonts}
\usepackage{epsfig}
\usepackage{graphicx}
\usepackage{natbib}
\usepackage{subfigure}

\title[Doppler Imaging in Sun-as-a-star observations]{New aspects of Doppler imaging in Sun-as-a-star observations}

\author[A.M. Broomhall et al.]{A.~M. Broomhall$^1$\thanks{E-mail:
amb@bison.ph.bham.ac.uk}, W.~J. Chaplin$^1$, Y. Elsworth$^1$, R. New$^2$\\
$^1$School of Physics and Astronomy, University of Birmingham,
Edgbaston, Birmingham B15 2TT\\$^2$Faculty of Arts, Computing,
Engineering and Sciences, Sheffield Hallam University, Sheffield S1
1WB}
\date{}
\begin{document}

\maketitle
\begin{abstract}
Birmingham Solar Oscillations Network (BiSON) instruments use
resonant scattering spectrometers to make unresolved Doppler
velocity observations of the Sun. Unresolved measurements are not
homogenous across the solar disc and so the observed data do not
represent a uniform average over the entire surface. The influence
on the inhomogeneity of the solar rotation and limb darkening has
been considered previously \citep{Brookes1978} and is well
understood. Here we consider a further effect that originates from
the instrumentation itself. The intensity of light observed from a
particular region on the solar disc is dependent on the distance
between that region on the image of the solar disc formed in the
instrument and the detector. The majority of BiSON instruments have
two detectors positioned on opposite sides of the image of the solar
disc and the observations made by each detector are weighted towards
differing regions of the disc. Therefore the visibility and
amplitudes of the solar oscillations and the realization of the
solar noise observed by each detector will differ. We find that the
modelled bias is sensitive to many different parameters such as the
width of solar absorption lines, the strength of the magnetic field
in the resonant scattering spectrometer, the orientation of the
Sun's rotation axis, the size of the image observed by the
instrument and the optical depth in the vapour cell. We find that
the modelled results best match the observations when the optical
depth at the centre of the vapour cell is 0.55. The inhomogeneous
weighting means that a `velocity offset' is introduced into
unresolved Doppler velocity observations of the Sun, which varies
with time, and so has an impact on the long-term stability of the
observations.
\end{abstract}

\begin{keywords}methods: observational, Sun: helioseismology, Sun:
oscillations
\end{keywords}

\section{Introduction}
Birmingham Solar Oscillations Network (BiSON) instruments use
resonant scattering spectrometers to make unresolved observations of
the Sun \citep{Brookes1978a}. This involves determining the intensity of light scattered
by a potassium vapour into the blue and red wings of a solar
Fraunhofer line. If the solar Fraunhofer line is Doppler shifted the
intensities observed on the blue and red wings will not be equal and
this difference allows the line-of-sight velocity between the Sun
and the instrument to be determined. Since BiSON instruments make
unresolved observations the intensity recorded in each wing is
integrated over the entire solar disc. \citet{Brookes1978} showed
that, because of the solar rotation and limb darkening, unresolved observations are not homogenous across the solar
disc and so the observed data do not represent a uniform average
over the entire surface.

Here we consider a further influence on the Doppler imaging that
originates from the instrumentation itself. The key component of
each BiSON instrument is a cell containing potassium vapour. The
intensity of light observed from a particular region on the solar
disc is dependent on the position of the detector with respect to
the image of the solar disc that is formed in the cell. The majority
of BiSON instruments have two detectors, called port and starboard,
which are positioned on opposing sides of the observed solar image.
In this paper we demonstrate that observations made by each detector
are weighted towards differing regions of the solar disc. This has
consequences for the visibility of the modes and the coherency of
the data observed from the two detectors.

The layout of the rest of this paper is as follows. We begin, in
Section \ref{section[BiSON instrumentation]}, by giving a brief
description of the BiSON instrumentation. Then, in Section
\ref{section[model limb and rotation]}, we use this information to
model the bias across the solar disc that is seen in unresolved
Doppler velocity observations. We have modelled the effect of the
solar rotation and limb darkening, both of which have been
considered previously \citep{Brookes1978}, and we show that the
observed bias is also dependent on the position of the detector with
respect to the image of the Sun captured by an instrument. The
line-of-sight velocity between the Sun and an instrument is
constantly varying due to the Earth's orbital and spin velocities
and in Section \ref{section[wmv]} we determine the effect of this
variation on the raw measured data. We also investigate how
sensitive the model results are to various input parameters, such as
the width of the solar absorption line, the width and separation of
the instrumental absorption lines, the position angle of the Sun's
rotation axis, the size of the image seen by the instrument and the
optical depth of the vapour in the instrument's cell. The total
observed line-of-sight velocity between the Sun and an instrument is
determined by calculating a ratio of intensities. Comparisons are
made between the observed and modelled ratios to obtain estimates of
the optical depth in the vapour cell and the size of the image of
the Sun observed by an instrument. We then use the observed and
modelled ratios to determine a `velocity offset' that is present in
the BiSON data (Section \ref{section[observed velocity offset]}). We
aim to model this velocity offset as it affects the long-term
stability of the BiSON observations and introduces noise into the
data. The main results of this paper and consequences of the
observed Doppler imaging are discussed in Section
\ref{section[discussion]}.

\section{BiSON instrumentation}\label{section[BiSON
instrumentation]} BiSON observations are made using resonant
scattering spectrometers, which determine the magnitude of Doppler
velocity shifts in the solar potassium Fraunhofer line at
$769.9\,\rm nm$ \citep{Brookes1978a}. Potassium vapour is enclosed
in a vapour cell and placed in a magnetic field, typically of
strength $0.18\,\rm T$, which corresponds to a separation in the
left- and right-circularly polarized Zeeman components of the
laboratory emission line of $0.013\,\rm nm$. This is equivalent to a
Doppler shift of $5200\,\rm m\,s^{-1}$. The right-circularly
polarized Zeeman component is shifted to a longer wavelength and so
will be referred to as the red instrumental component in the
remainder of this paper. The left-circularly polarized Zeeman
component is shifted to a shorter wavelength and so will be referred
to as the blue instrumental component. The two shifted instrumental
components are located on opposing wings of the potassium solar
Fraunhofer line.

In addition to any oscillation velocities that might be observed,
the total observed line-of-sight velocity between the Sun and the
instrument, $v_{\textrm{\scriptsize{obs}}}$, contains several
components: the velocity due to the Earth's orbit around the Sun,
$v_{\textrm{\scriptsize{orb}}}$, the velocity due to the Earth's
spin on its own axis, $v_{\textrm{\scriptsize{spin}}}$, the
gravitational redshift, $v_{\textrm{\scriptsize{grs}}}$, and various
other offsets, collectively known as
$v_{\textrm{\scriptsize{other}}}$, which include the Doppler imaging
offset that will be discussed in this paper and any offsets due to
noise. The BiSON group uses the JPL ephemeris look-up tables, which
are available through the JPL Horizons system, to determine the sum
of the line-of-sight velocity due to $v_{\textrm{\scriptsize{orb}}}$
and $v_{\textrm{\scriptsize{spin}}}$. We define the `station
velocity', $v_{\textrm{\scriptsize{st}}}$, as the sum of these two
velocities, i.e.,
\begin{equation}\label{equation[station velocity]}
    v_{\textrm{\scriptsize{st}}}=v_{\textrm{\scriptsize{orb}}}+v_{\textrm{\scriptsize{spin}}}.
\end{equation}

To determine the observed Doppler velocity of the Sun with respect
to the instrument, $v_{\textrm{\scriptsize{obs}}}$, the following
ratio is calculated from the raw observations:
\begin{equation}\label{equation[ratio1]}
    R=\frac{I_b-I_r}{I_b+I_r},
\end{equation}
where $I_b$ and $I_r$ are the observed strengths of the resonantly
scattered signal on the blue and red sides of the solar line,
respectively. At small line-of-sight velocities the blue and red
instrumental components are positioned on the parts of the solar
absorption line with the steepest gradient, making the measurements
very sensitive to small changes in line-of-sight velocity. However
at large line-of-sight velocities the instrumental components are
positioned on shallower portions of the solar absorption line, thus
reducing the sensitivity of the measurements. In other words, the
sensitivity is dependent on the magnitude of the line-of-sight
velocity between the Sun and the instrument.

We model the relationship between the station velocity,
$v_{\textrm{\scriptsize{st}}}$, and the ratio, $R$, with the
following equation:
\begin{equation}\label{equation[ratio2]}
    R(v_{\textrm{\scriptsize{st}}})=
    \sum^N_{i=0}a_iv_{\textrm{\scriptsize{st}}}^{\,i},
\end{equation}
where $N$ is the order of the polynomial and $a_i$ are the
coefficients of the polynomial. The order of the polynomial must be
chosen carefully. If there the number of terms included is too small
there are insufficient degrees of freedom in the model to follow the
change in sensitivity throughout the day. However, if too many terms
are included in the polynomial the fit becomes unstable and the
signal from the oscillations can be lost when the series is
subtracted from the measured values. Currently, when the BiSON data
are processed, a cubic $(N=3)$ polynomial is used
\citep{Elsworth1995}. A best fit algorithm is used to determine the
coefficients, $a_i$, in equation \ref{equation[ratio2]} and the
gravitational redshift, $v_{\textrm{\scriptsize{grs}}}$, on a
day-by-day basis. In the remainder of this paper, when modelling the
observed Doppler imaging, we take
$v_\textrm{\scriptsize{grs}}=632\,\rm m\,s^{-1}$. This is marginally
higher than the offset measured by the instruments, as it does not
account for the observed convective blue shift, but it is a good
approximation. The modelled ratio, given by equation
\ref{equation[ratio2]}, is subtracted from the observed ratio, as
defined by equation \ref{equation[ratio1]}, to leave a residual from
which the oscillation velocities can be determined.

We now use this information about the manner in which BiSON
instruments measure velocity shifts to determine the observed bias
across the solar disc.

\section{Modelling the bias across the disc}\label{section[model limb and
rotation]}

\begin{figure}
  \centering
  \subfigure{\includegraphics[width=0.3\textwidth, clip]{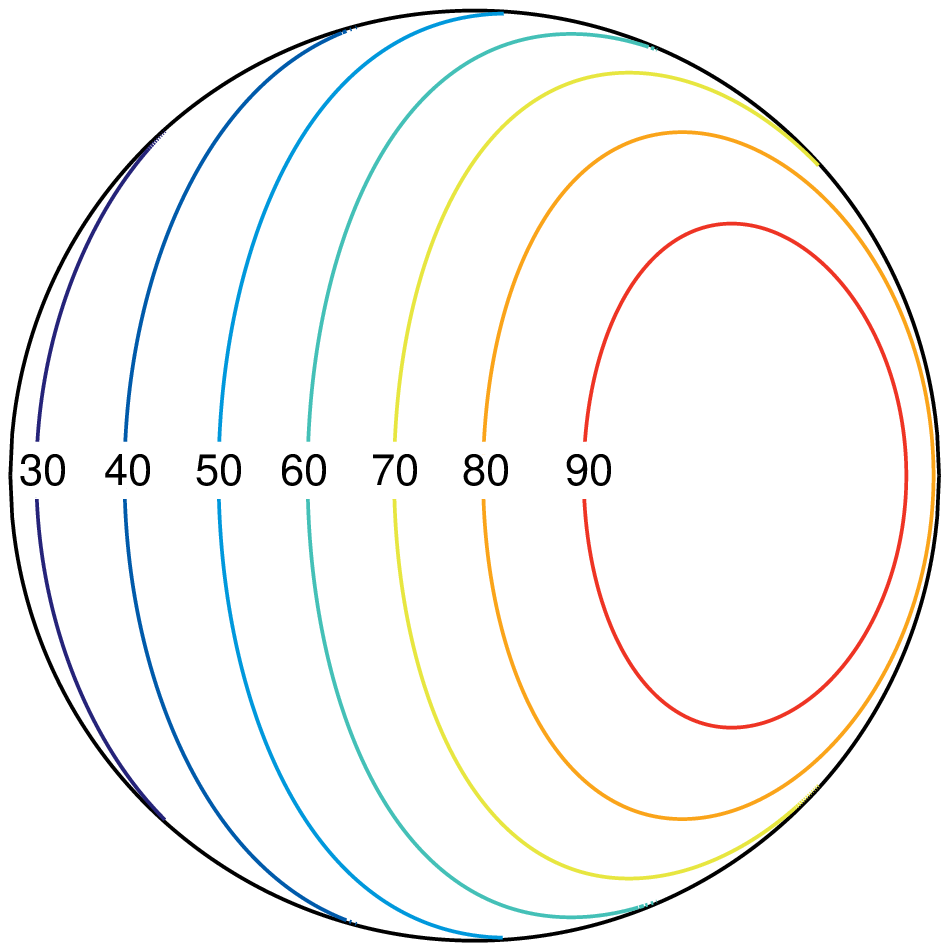}}
  \hspace{1cm}
  \subfigure{\includegraphics[width=0.3\textwidth, clip]{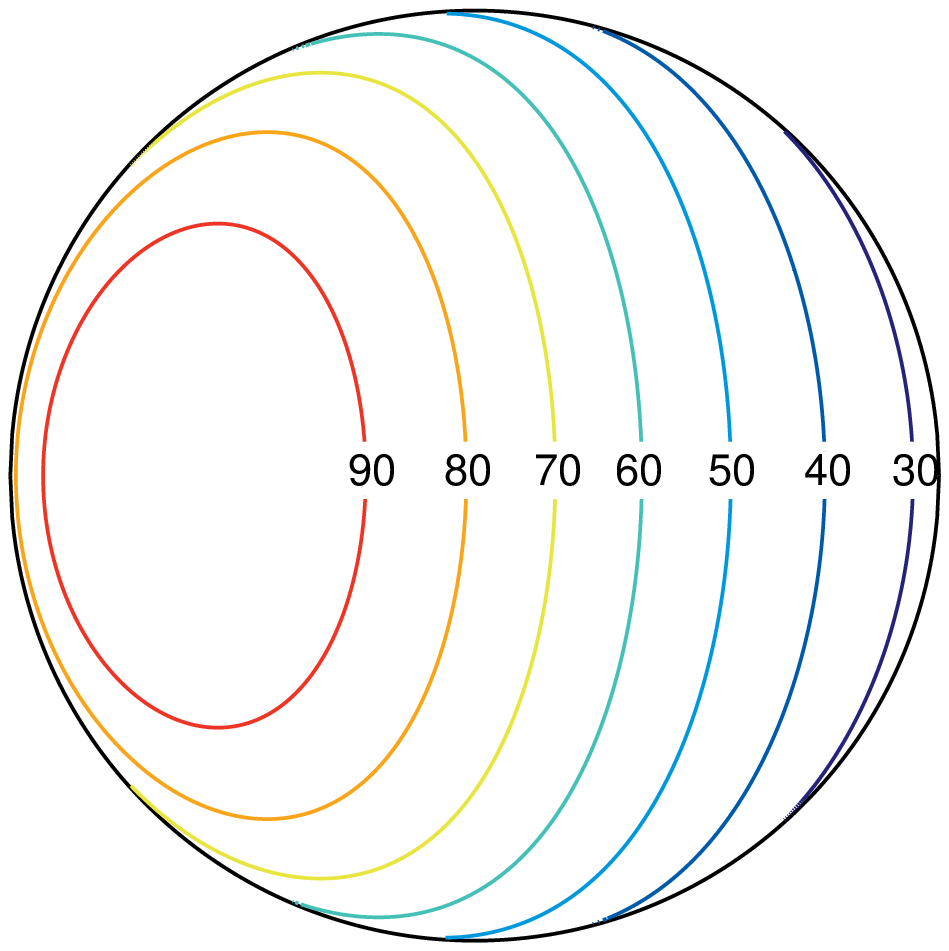}}\\
  \caption[Weighting of the solar disc when limb darkening and the solar rotation
  are considered and
  $v_{\textrm{\scriptsize{los}}}=0\,\rm m\,s^{-1}$.]
  {The weighting of the solar disc due to limb darkening and the solar rotation.
  The top panel shows the results for the
  instrumental blue wing, while the bottom panel shows the results
  for the instrumental red wing.
  The station velocity,
  $v_{\textrm{\scriptsize{st}}}$, is $0\,\rm m\,s^{-1}$.}
  \label{figure[weighting limb and vel]}
\end{figure}

Figure \ref{figure[weighting limb and vel]} shows the weighting of
the solar disc caused by limb darkening and the solar rotation. This
work has been performed previously by \citet{Brookes1978} but is
included here so that the weightings can be compared with those
observed when we take into account instrumental effects (in Section
\ref{section[detector weighting]}). The weightings shown in Figure
\ref{figure[weighting limb and vel]} are scaled relative to the
region that contributes most to the unresolved observations. So a
weighting of 100 corresponds to the region that contributes most to
the unresolved observations and a weighting of 50 corresponds to a
region that contributes half of the maximum. Contours with
weightings from 30 to 90 have been plotted.

The Sun's rotation axis was taken to be vertical and the period of
rotation was taken as the constant value of 25 days, which
corresponds to the observed rotation period of the material at the
solar equator. The Sun actually exhibits differential rotation, such
that the rotation period of material at the poles is approximately
36 days and so assuming a uniform rotation at all latitudes is
rather simplistic. However, increasing the rotation period to 36
days has a negligible effect on the calculated bias, implying the
results would not be altered significantly by the inclusion of
differential rotation. The rotation of the Sun is in an
anti-clockwise direction if viewed from above, looking down on the
north pole (north is upwards in Figure \ref{figure[weighting limb
and vel]}). Therefore, in Figure \ref{figure[weighting limb and
vel]}, the left solar limb is approaching an observer on Earth,
while the right solar limb is receding.

As can be seen in Figure \ref{figure[weighting limb and vel]}
observations made with the red instrumental wing are weighted most
heavily towards the approaching limb of the solar disc, while
observations made with the blue instrumental wing are most heavily
weighted towards the receding solar limb. This is expected as BiSON
spectrometers observe a narrow absorption line and so, for example,
the blue instrumental wing observes a larger intensity of light from
regions of the solar disc that are red shifted i.e. the receding
limb of the solar disc.

It should be noted that the results shown in Figure
\ref{figure[weighting limb and vel]} have been calculated when
$v_{\textrm{\scriptsize{st}}}+v_{\textrm{\scriptsize{grs}}}=0\,\rm
m\,s^{-1}$. However, the observed bias continually changes as
$v_{\textrm{\scriptsize{st}}}$ varies. In Section \ref{section[wmv]}
we examine how the variations in $v_{\textrm{\scriptsize{st}}}$
alter the observed bias. However, first we add to this model of the
observed inhomogeneity by considering the effect the position of the
detector has on the observed bias.

\subsection{Weighting of the solar image due to the position of the
detector}\label{section[detector weighting]}

\begin{figure}
  \centering
  \includegraphics[width=0.35\textwidth, clip]{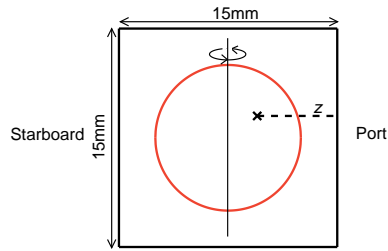}\\
  \caption[A schematic of a vapour cell.]{A schematic of a vapour cell. The vapour cell has dimensions $15\,\rm mm\times 15\,\rm
  mm$. At the centre of the vapour cell is the observed image of the
  Sun (red circle). The port and starboard detectors are positioned
  on either side of the vapour cell. Here we take the optical path
  of light to the detector to be horizontal. In this figure the optical path is shown by the dashed line and $z$ is
  the distance travelled by light through the vapour. Also shown are
  the axis and direction of rotation. Note that the actual orientation
  of the rotation axis is dependent on season (see Section \ref{section[observed velocity offset]}).}\label{figure[vapour_cell]}
\end{figure}

Consider a vertical slice through the vapour cell (see Figure
\ref{figure[vapour_cell]}), which is a square of side $15\,\rm mm$
and the port and starboard detectors are positioned on the right-
and left-hand sides of the vapour cell respectively. The intensity
of light detected from a particular region on the solar disc by the
port detector depends on the distance between that region on the
image of the Sun in the vapour cell and the inside wall of the
vapour cell on the side of the port detector ($z$ in Figure
\ref{figure[vapour_cell]}). Here we have only considered the
perpendicular distance between the region on the image and the wall
of the vapour cell. This is not strictly true in a real instrument,
which has a system of lenses between the image and the detector, but
it does provide a good first-order approximation.

The intensity observed by the detector from a given region on the
image of the Sun, $I$, is determined by the optical depth of the
vapour in the cell along the path, $\tau$, and is given by
\begin{equation}\label{equation[scattered intensity]}
    I = I_0\textrm{e}^{-\tau},
\end{equation}
where $I_0$ is the intensity of light received from the Sun. The
optical depth is given by
\begin{equation}\label{equation[scattered intensity]}
    \tau = Kz,
\end{equation}
where $K$ is the extinction coefficient and $z$ is the horizontal
path length (see Figure \ref{figure[vapour_cell]}). We have assumed
that $K$ is constant throughout the vapour cell and so as $z$ increases $I$ decreases. It should be noted
here that this model is based on the assumption that only single
scattering is significant. We define $\tau_0$ as the optical depth at the centre of the cell, when $z=7.5\,\rm mm$.

We now consider the effect on the weighting of unresolved solar
observations of the detector position alone and then we combine
this effect with the previously considered results for limb
darkening and solar rotation.

\subsubsection{Weighting of the solar image observed by the port and
starboard detectors}
\begin{figure}
  \centering
  \subfigure[Starboard detector]{\includegraphics[width=0.3\textwidth,
  clip]{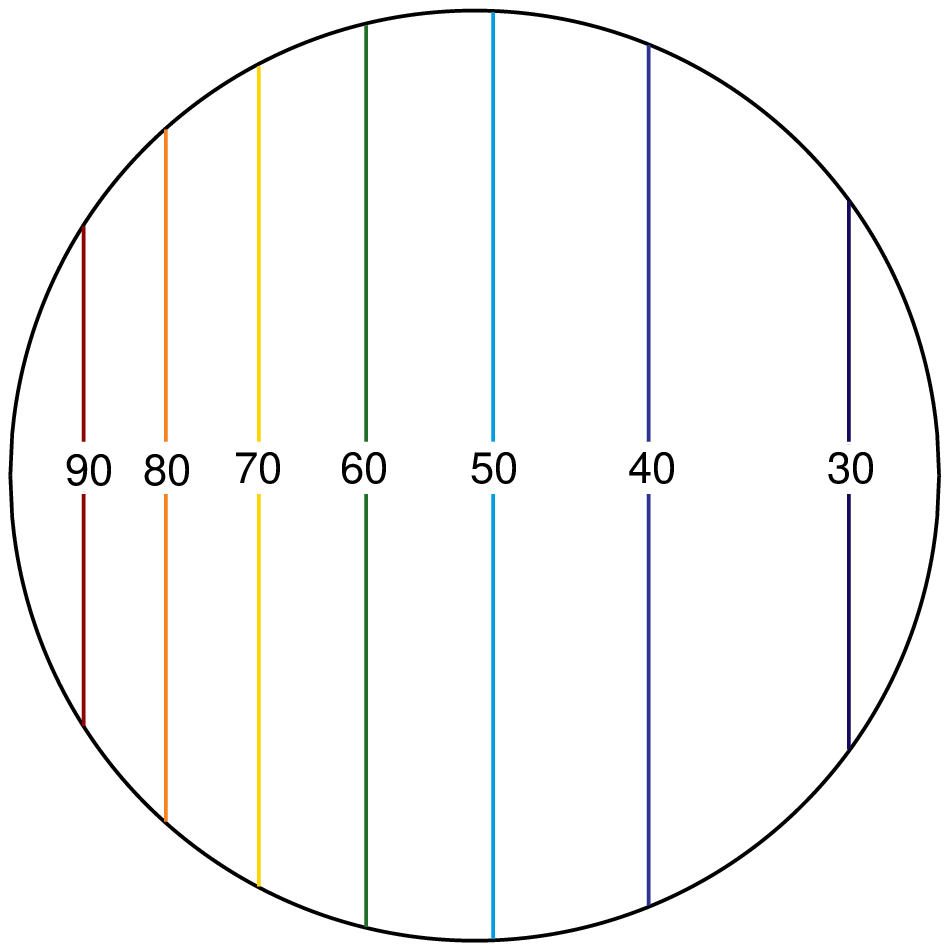}}\\
  \subfigure[Port
  detector]{\includegraphics[width=0.3\textwidth, clip]{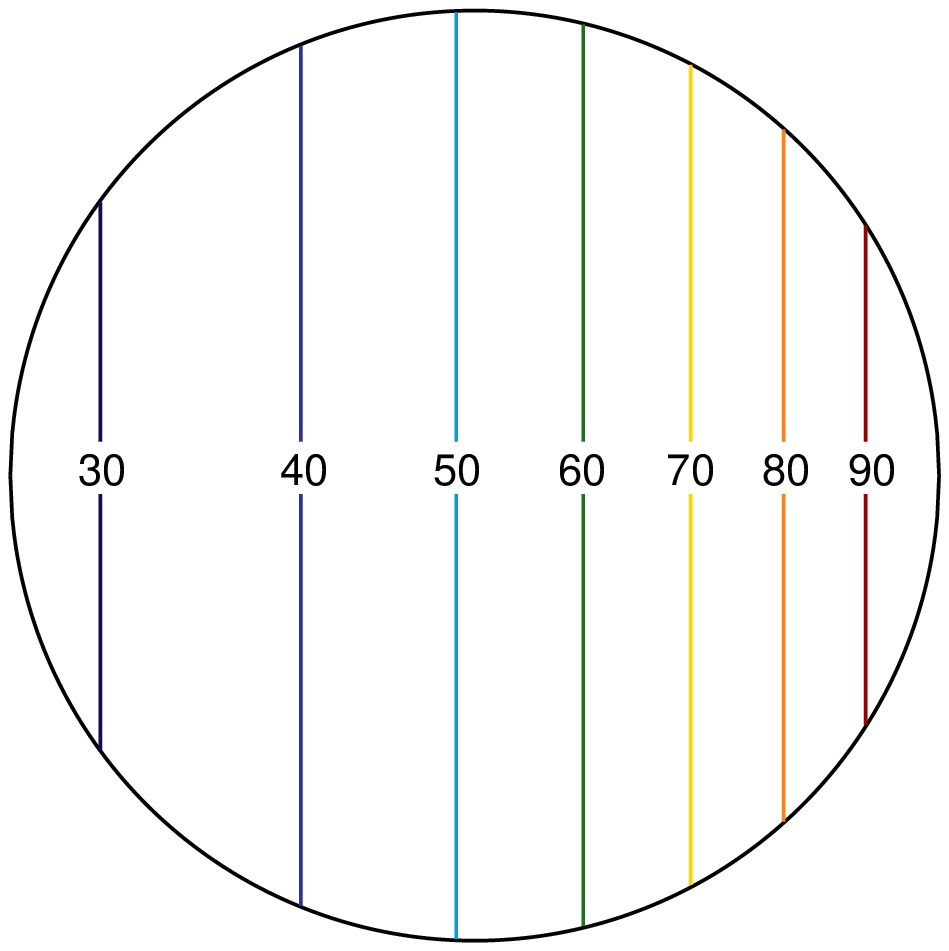}}\\
  \caption{The weighting across the solar disc caused by the positioning of
  the detector.}\label{figure[contour p and s]}
\end{figure}

Figure \ref{figure[contour p and s]} shows the results of
considering just the effect of observing on different sides of the
vapour cell. As in Figure \ref{figure[weighting limb and vel]} the
most heavily weighted region is given a weighting of 100. The other
values of the other contours are then determined as a percentage of
the maximum i.e. a contour of 50 means the region contributes half
the amount contributed by a region with a contour of 100. Contours
with values from 30 to 90 have been plotted. We have assumed that
the vapour has an optical depth at the centre of the cell, $\tau_0$,
of unity and so $K=133.3\,\rm m^{-1}$. In later sections we change
the value of $\tau_0$ (see Sections \ref{subsection[tau change]},
\ref{subsection[find tau r]} and \ref{section[observed velocity
offset]}). We gave the image of the solar disc in the vapour cell a
radius of $5\,\rm mm$ as this is known to be the approximate size
the solar image in a BiSON vapour cell. In the remainder of this
paper we assume that the centre of the image of the solar disc
coincides with the centre of the vapour cell. This is not strictly
true for the real BiSON instruments but a shift away from the centre
of the cell will not alter the relative weighting across the disc.

The observations are weighted most heavily towards the side of the
solar disc that is closest to the detector. Thus, the starboard
observations are weighted towards the approaching limb of the Sun,
while the port observations are weighted towards the receding limb.
As we have only considered the horizontal distance between a pixel
on the solar image and the edge of the vapour cell the contours of
constant weighting are vertical lines. It should be noted that the
effect of the detector's position on the observed bias is
independent of which instrumental absorption wing is considered, and
the station velocity.

We now move on to amalgamate the effects of the Sun's rotation, limb
darkening and the position of the detector.

\begin{figure*}
  \centering
  \subfigure[Blue wing, starboard
  detector]{\includegraphics[width=0.3\textwidth, clip]{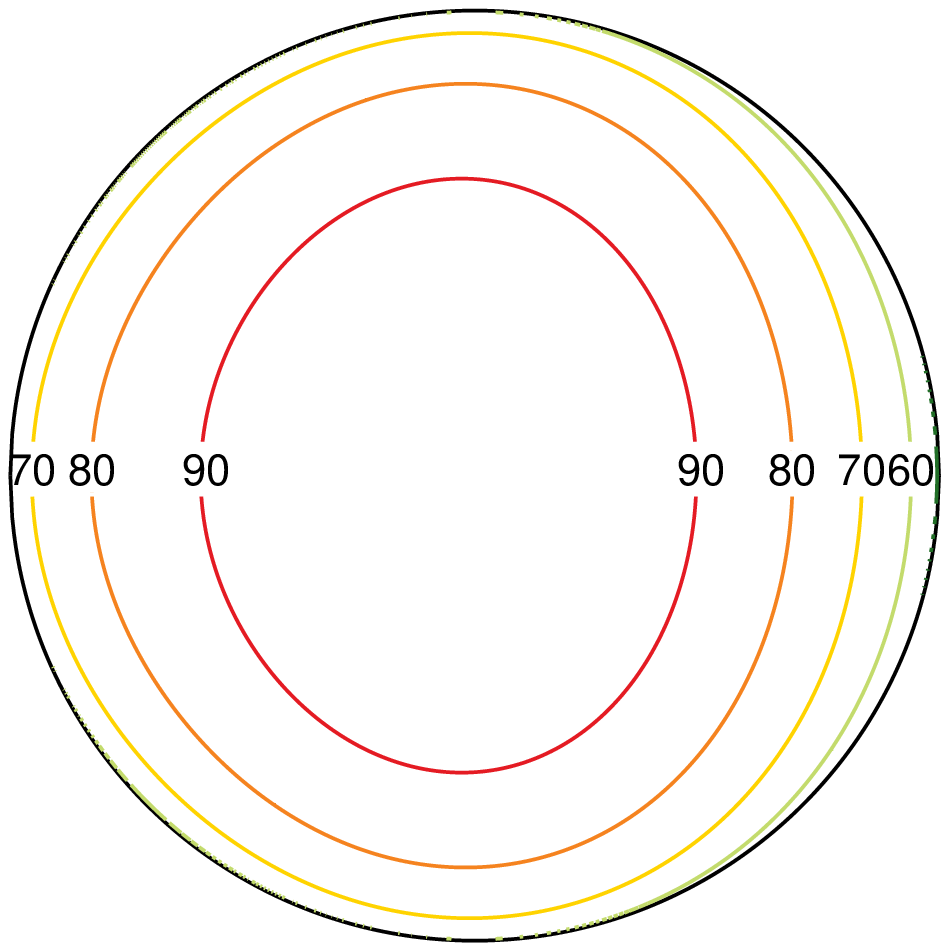}}
  \hspace{1cm}
  \subfigure[Red wing, starboard detector]{\includegraphics[width=0.3\textwidth, clip]{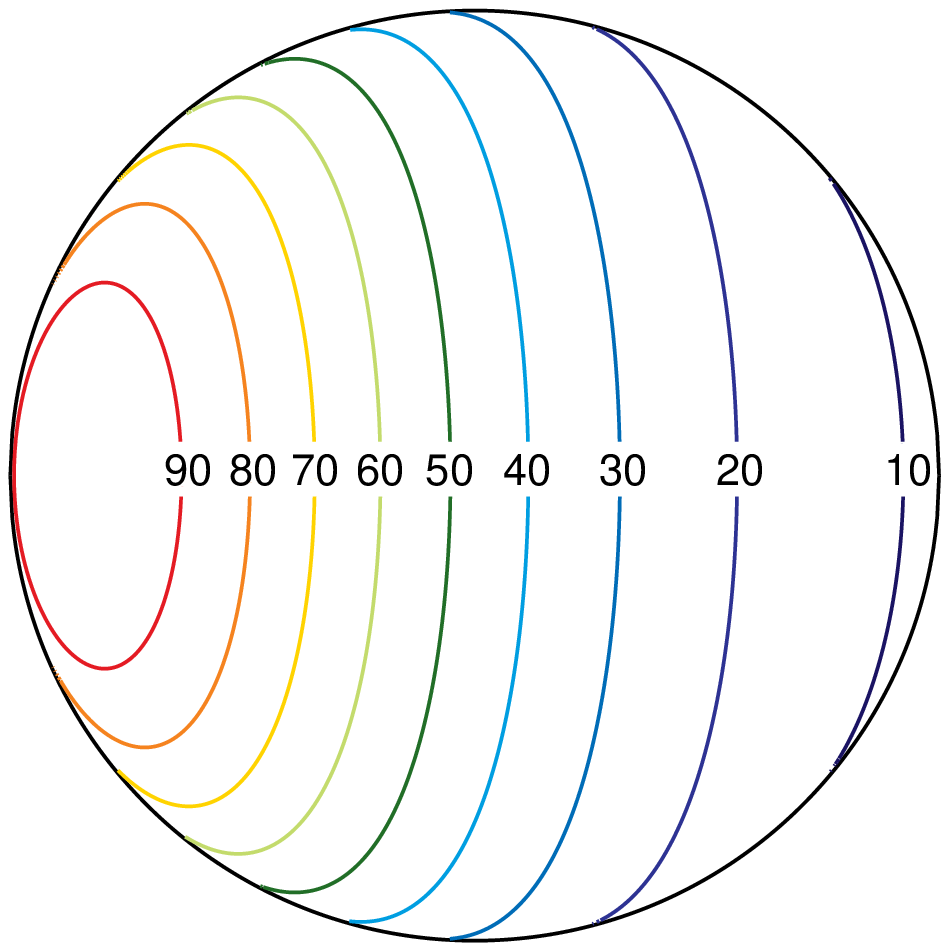}}\\
  \vspace{0.5cm}
  \subfigure[Blue wing, port
  detector]{\includegraphics[width=0.3\textwidth, clip]{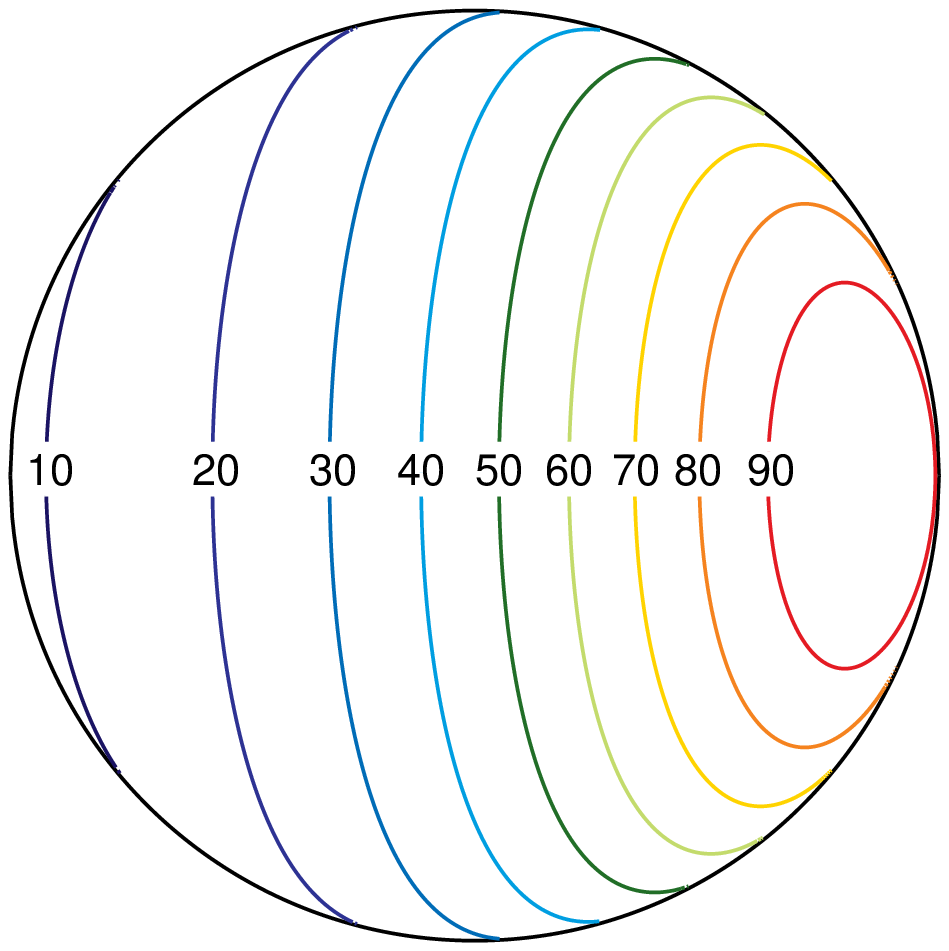}}
  \hspace{1cm}
  \subfigure[Red wing, port detector]{\includegraphics[width=0.3\textwidth, clip]{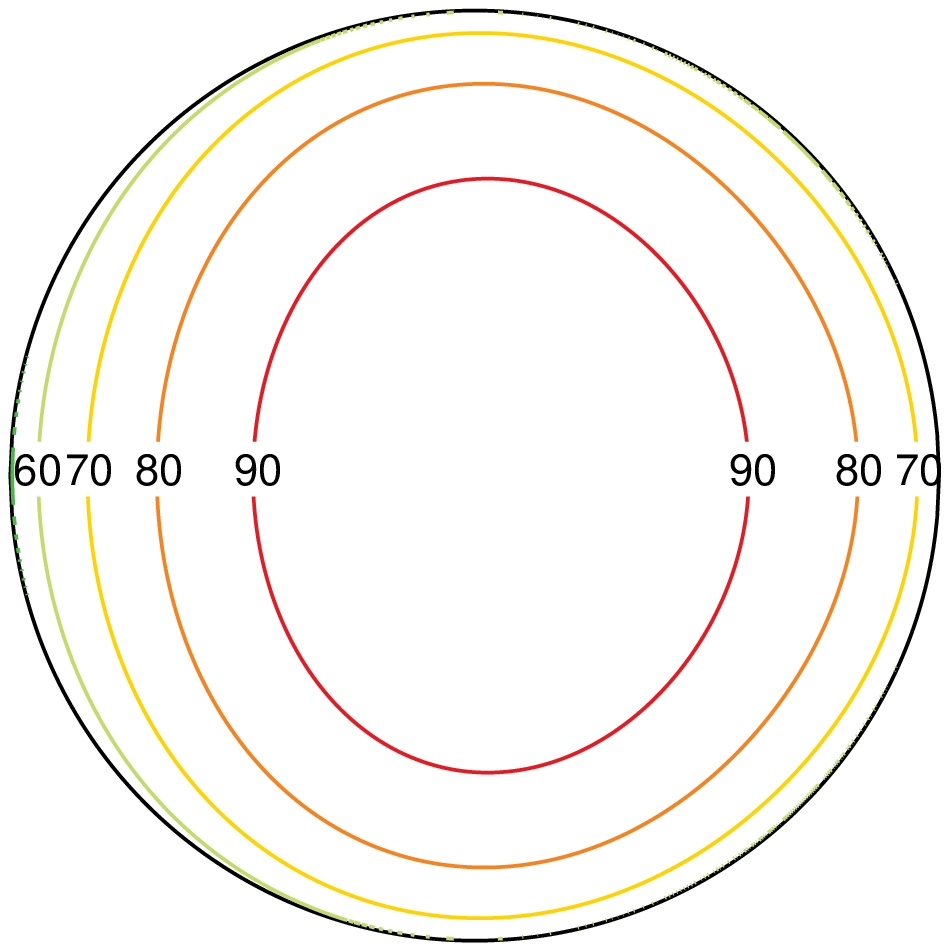}}\\
  \caption{Contour maps showing the weighting of the solar disc that is observed when the position
  of the detector is considered as well as the limb darkening and
  the effect of the solar rotation. The radius of the image in the vapour
  cell is $5\,\rm mm$, $v_{\textrm{\scriptsize{st}}}+v_{\textrm{\scriptsize{grs}}}=0\,\rm m\,s^{-1}$ and the optical
  depth at the centre of the cell is unity.}\label{figure[all contour vlos0]}
\end{figure*}

\subsection{Weighting of the solar disc due to the combined effect
of the solar rotation, limb darkening and the position of the
detector}

Figure \ref{figure[all contour vlos0]} shows the weighting of the
solar disc in Sun-as-a-star Doppler velocity observations when the
effects of the solar rotation, limb darkening and the position of
the detector are considered and
$v_{\textrm{\scriptsize{st}}}+v_{\textrm{\scriptsize{grs}}}=0\,\rm
m\,s^{-1}$. The values of the contours are defined in the same way
as for Figures \ref{figure[weighting limb and vel]} and
\ref{figure[contour p and s]}. Comparison between Figure
\ref{figure[all contour vlos0]} and Figure \ref{figure[weighting
limb and vel]} implies that the position of the detector is very
important to the observed bias across the solar disc. Before the
instrumental effect was considered the blue wing observations were
weighted towards the receding limb of the Sun (the right-hand limb
in Figure \ref{figure[weighting limb and vel]}). When the Sun is
observed by the starboard detector the results are shifted towards
the approaching solar limb (the left-hand limb in Figure
\ref{figure[all contour vlos0]}). The balance between these two
effects results in the observations being weighted most heavily
towards the centre of the disc. Similar reasoning can be used to
explain why the red, port observations are also weighted towards the
centre of the solar disc. Closer inspection reveals that the blue
starboard observations are weighted very slightly towards the
approaching limb of the Sun (left-hand limb in Figure
\ref{figure[all contour vlos0]}), while the red port observations
are weighted slightly towards the receding solar limb. The results
observed in Figure \ref{figure[all contour vlos0]} therefore imply
that the dominant effect in determining the weighting across the
solar disc is that of the position of the detector.

When
$v_{\textrm{\scriptsize{st}}}+v_{\textrm{\scriptsize{grs}}}=0\,\rm
m\,s^{-1}$ the red, starboard map is the mirror image of the blue,
port map and the red, port map is the mirror image of the blue,
starboard map. As $v_{\textrm{\scriptsize{st}}}$ increases the
weighting patterns are shifted towards the approaching limb of the
Sun and so the results are no longer symmetric.

When determining $I_r$ and $I_b$, and consequently the observed
Doppler velocity, $v_{\textrm{\scriptsize{obs}}}$, we sum the
intensities observed from each region on the solar disc. Therefore,
Figure \ref{figure[all contour vlos0]} implies that the calculated
observed velocity does not represent a uniform average across the
whole disc. We have now built up an image of the weighting of the
Sun that accounts for the solar rotation, limb darkening and the
position of the detector. We move on to investigate how the bias
varies with station velocity. We also consider the sensitivity of
the inhomogeneity to various inputs involved in the model
calculations.

\section{The sensitivity of the modelled ratio of the input parameters}\label{section[wmv]}

The ratio, $R$, which is given by equation \ref{equation[ratio1]},
is the raw datum produced from BiSON observations. Using the model
of the unresolved Doppler velocity observations described in Section
\ref{section[model limb and rotation]} this ratio can be determined
for different values of $v_{\textrm{\scriptsize{st}}}$.

\begin{figure}
  \centering
  \includegraphics[width=0.35\textwidth, clip]{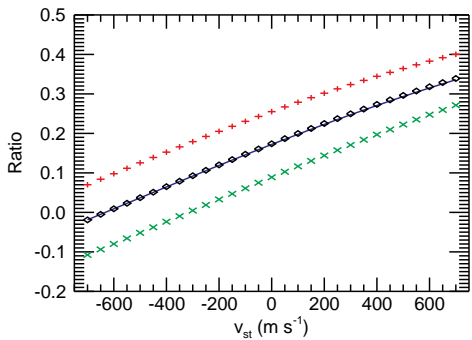}\\
  \caption
  {The modelled ratio as a function of station
  velocity. The black diamonds
  represent the results when only the effect of rotation and limb
  darkening are considered. The green crosses represent the starboard observations
  and the red plus signs represent the port observations.The solid blue line
  represents the mean of the port and starboard ratios.}
  \label{figure[blue-red weighted vel]}
\end{figure}

To determine the sensitivity of the ratio to various input
parameters we have calculated the ratio over a range of station
velocities and then varied the input parameters one by one. However,
we began by determining the ratio when the bias caused by the solar
rotation and the limb darkening only were included in the
calculations and the results are shown in Figure
\ref{figure[blue-red weighted vel]} (black diamonds). As expected,
when the sum of $v_{\textrm{\scriptsize{st}}}$ and
$v_{\textrm{\scriptsize{grs}}}$ is zero (i.e.
$v_{\textrm{\scriptsize{st}}}=-632\,\rm m\,s^{-1}$) the blue and red
wing observations are symmetric and so the ratio is zero. The ratio
increases as $v_{\textrm{\scriptsize{st}}}$ increases, however, the
relationship between $v_{\textrm{\scriptsize{st}}}$ and $R$ is not
linear as the observations become biased towards different portions
of the solar absorption line and different regions on the solar
surface.

Also plotted in Figure \ref{figure[blue-red weighted vel]} are the
ratios observed by the port and starboard detectors. In the
calculations the radius of the image of the Sun in the vapour cell
was $5\,\rm mm$ and the optical depth, $\tau_0$, was unity at the
centre of the cell. The results show that the ratio is dependent on
which detector is considered. The observations made by the port
detector are red shifted, while the observations made by the
starboard detector are blue shifted. Usually, when dealing with real
BiSON data, the mean of the port and starboard ratios is determined.
This mean ratio (the solid blue line in Figure \ref{figure[blue-red
weighted vel]}) is not absolutely identical to the ratio calculated
without considering the position of the detector (the black diamonds
in Figure \ref{figure[blue-red weighted vel]}).The difference
between the mean ratio and the detector-independent ratio increases
with $v_{\textrm{\scriptsize{st}}}$ and is of the order of 1 per
cent.

We move on to determine the sensitivity of the calculations to the
assumed values of various instrumental parameters.

\subsection{Sensitivity of the calculations to the width of the
solar absorption line}

\begin{figure}
  \centering
  \includegraphics[width=0.35\textwidth, clip]{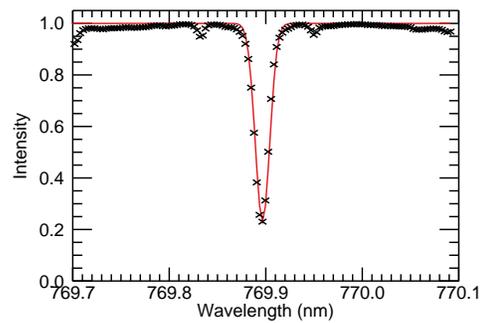}\\
  \caption{The Fraunhofer line observed by Themis when looking at the centre
  of the solar disc, where the line-of-sight velocity due to the solar
  rotation is zero (black crosses). Also plotted is a Gaussian approximation to this line (solid red line).}
  \label{figure[unshifted line]}
\end{figure}

\begin{figure}
\centering
  \includegraphics[width=0.35\textwidth, clip]{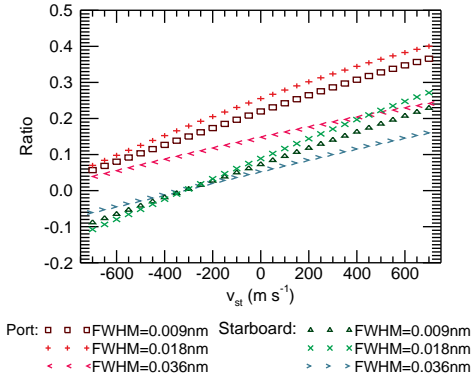}\\
  \caption{Sensitivity of the modelled ratio to the FWHM of
  the solar absorption line. The different symbols
  represent the results calculated using different values of the FWHM
  of the solar absorption line (see legend).}\label{figure[solar line width sensitivity]}
\end{figure}

One of the main input parameters in the calculations that determine
the ratio is the width of the solar absorption line. To determine
the weighting across the solar disc we consider the solar absorption
line observed from $\sim250,000$ pixels on the solar disc
individually. Each pixel is sufficiently small that the solar
absorption line observed from any one pixel will not be noticeably
broadened by the solar rotation, unlike a solar absorption line
observed from the whole Sun. We have used measurements made by the
Themis instrument on $\rm Iza\tilde{n}a$ \citep{Simoniello2008} to
estimate the width of a solar potassium absorption line that is
observed by a single pixel. Themis makes resolved observations of
the solar potassium absorption line from a small pixel at the centre
of the solar disc. Consequently, the solar absorption line observed
by Themis, which is shown in Figure \ref{figure[unshifted line]}, is
not broadened by the solar rotation and so can be used to model the
solar absorption line observed from an individual pixel on the Sun.
The scanned absorption profile observed by Themis is well
approximated by a Gaussian with a full width at half maximum (FWHM)
of $0.018\,\rm nm$ (see Figure \ref{figure[unshifted line]}). To
investigate the sensitivity of the ratio to the width of the solar
absorbtion line, calculations were performed when the FWHM of the
solar absorption line observed by one pixel was increased to
$0.036\,\rm nm$ and decreased to $0.009\,\rm nm$. The results are
shown in Figure \ref{figure[solar line width sensitivity]}.

The gradient of the variation in the ratio with
$v_{\textrm{\scriptsize{st}}}$ is very sensitive to the assumed FWHM
of the solar absorption line. It is, therefore, very important to
ensure that the assumed value of the width of the solar absorption
line is correct if the ratio is to be modelled accurately. Notice
that the gradient is steepest when the FWHM of the solar absorption
line is that of the Themis line $(0.018\,\rm nm)$. This is because
BiSON instruments have been set up so that the observations are as
sensitive as possible to any velocity changes. When the FWHM of the
solar line is $0.018\,\rm nm$ the instrument measures the intensity
of light on the steepest linear parts of the absorption line.
However, when the width of the solar line is increased the
instrument determines the intensities on the shallow base of the
solar line and so is less sensitive to changes in velocity.
Similarly, when the FWHM of the solar line is decreased observations
are made of the outer parts of the solar line, where the gradient is
also less steep and therefore the sensitivity is again decreased.

\subsection{Sensitivity of the calculations to the width and
separation of the instrumental absorption lines}

When calculating the effect of the Sun's rotation on the bias across
the solar disc we had to include various parameters that described
the instrumental absorption lines. We took the FWHM of these lines
to be $700\,\rm m\,s^{-1}$ (or $0.0018\,\rm nm$), based on the
Doppler broadening expected given the temperature of the vapour
cell. We took the separation between the blue and red instrumental
wings to be $5200\,\rm m\,s^{-1}$ (or $0.013\,\rm nm$), based on the
Zeeman splitting expected when the potassium vapour in the cell is
placed in a magnetic field of $0.18\,\rm T$. Here we explore how
sensitive the calculations are to variations in the input values of
these parameters.

We determined the ratio at different values of
$v_{\textrm{\scriptsize{st}}}$ when the widths of the instrumental
absorption lines were increased to $1400\,\rm m\,s^{-1}$ (or
$0.0036\,\rm nm$). Figure \ref{figure[instrument parameters
weighting]} indicates that doubling the width of the instrumental
absorption lines alters the calculated ratio by only a small amount.
The maximum change in the ratio, which occurs at large
$v_{\textrm{\scriptsize{st}}}$, is $\sim0.09$, and so this is not an
important effect.

\begin{figure}
\centering
  \includegraphics[width=0.32\textwidth, clip]{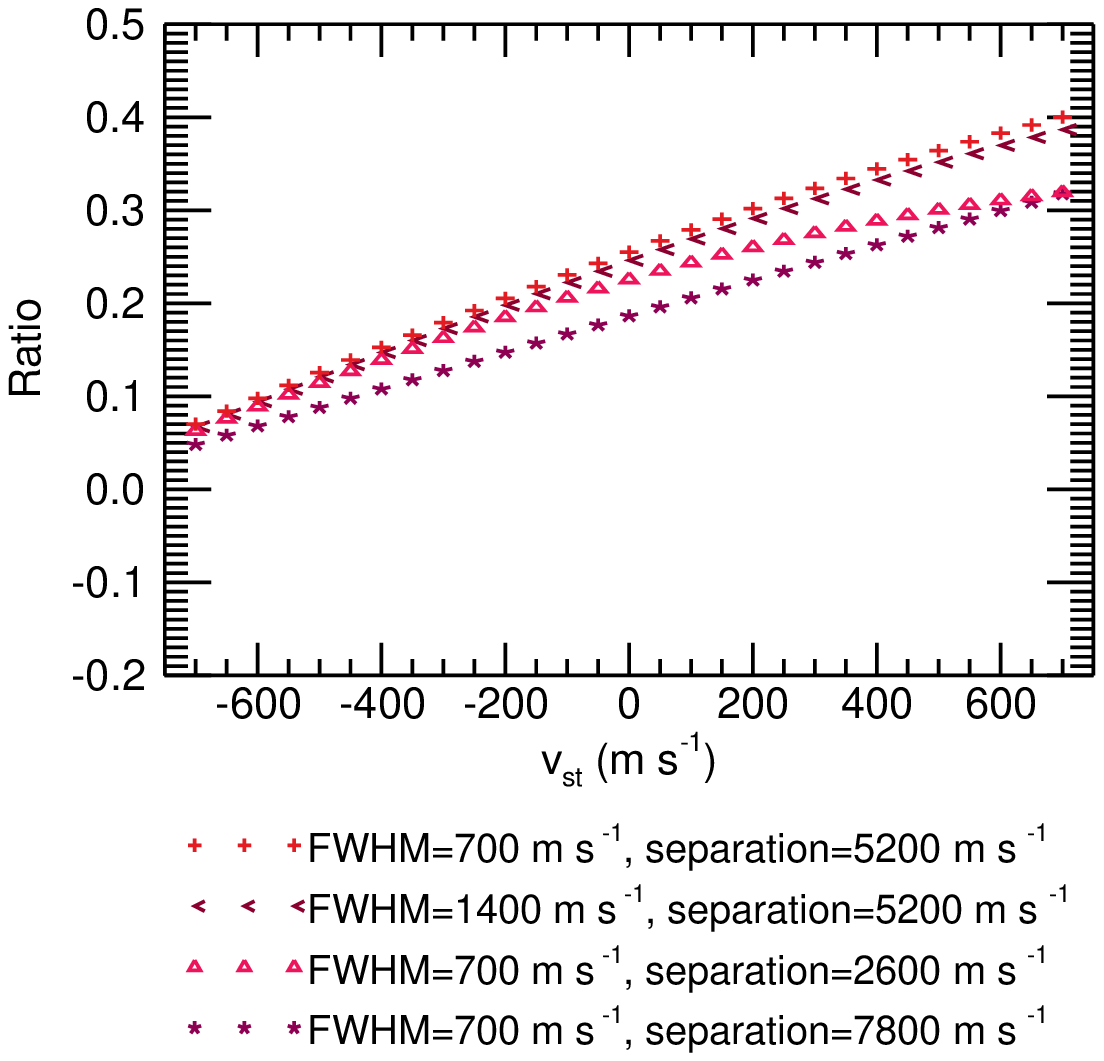}\\
  \vspace{0.5cm}
  \includegraphics[width=0.32\textwidth, clip]{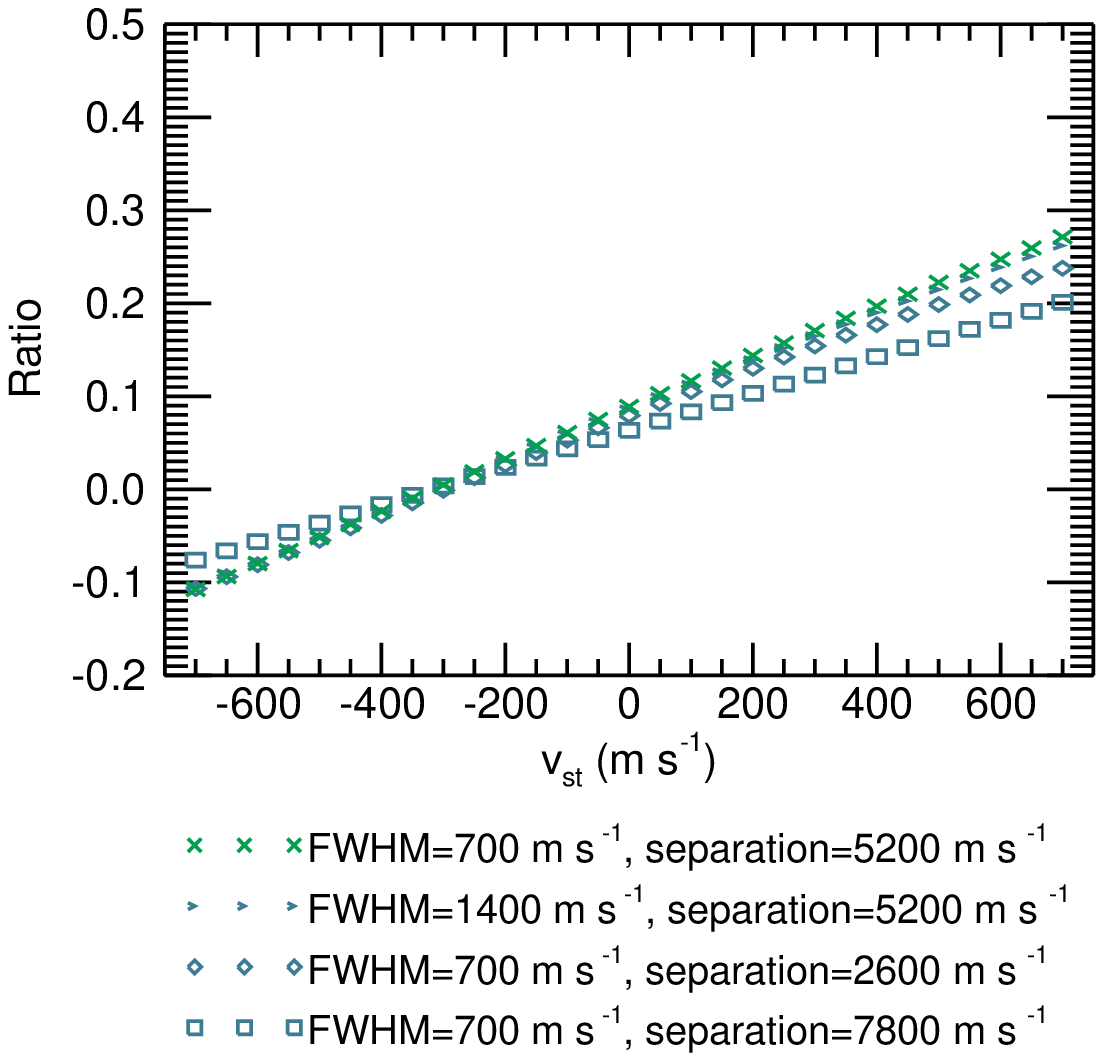}
  \caption{Sensitivity of the modelled ratio to the FWHM and separation of the instrumental absorption lines.
  The top panel shows the results for the port detector and the bottom panel
  shows the results for the starboard detector. Each symbol
  represents different values of the FWHM and separation of the
  instrumental absorption lines (see legends).}
  \label{figure[instrument parameters weighting]}
\end{figure}

Also plotted on Figure \ref{figure[instrument parameters weighting]}
are the results obtained when the separation of the instrumental
components was decreased to $2600\,\rm m\,s^{-1}$ (or $0.0067\,\rm
nm$) and increased to $7800\,\rm m\,s^{-1}$ (or $0.0200\,\rm nm$).
In reality this could be achieved by altering the strength of the
magnetic field in which the vapour cell is placed. In these
calculations the width of the instrumental absorption lines were
taken as $700\,\rm m\,s^{-1}$. Altering the separation of the
instrumental components affects the gradient of the variation of the
ratio with $v_{\textrm{\scriptsize{st}}}$. Therefore, it is
important that the separation of the blue and red instrumental
absorption lines is known to allow precise models of the ratio to be
made.

\subsection{Sensitivity of the calculations to the position angle of the Sun's rotation axis}

\begin{figure}
\centering
  \includegraphics[width=0.32\textwidth, clip]{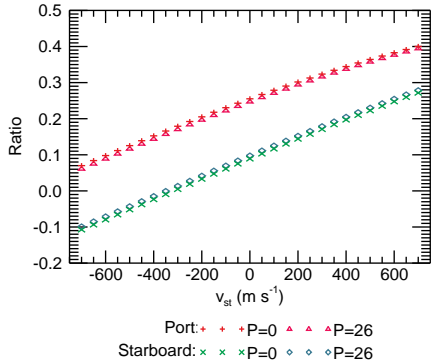}\\
  \caption{Sensitivity of the modelled ratio to the orientation of
  the Sun's rotation axis. The results are plotted for $P=0^\circ$
  and $P=26^\circ$ (see legend).}
  \label{figure[sensitivity P]}
\end{figure}

The position angle of the Sun's rotation axis, $P$, varies as Earth
orbits the Sun and can be tilted by more than $26^\circ$ from
vertical. The observed value of $P$ for any day can be found in a
standard almanac. The modelled ratio was determined when the
position angle of the Sun's rotation axis was $P=26^\circ$ and the
results are plotted in Figure \ref{figure[sensitivity P]}. Note that
the change in ratio is independent of the sign of $P$ i.e. the same
results were obtained when $P=-26^\circ$. Although only a small
effect, the determined ratios are sensitive to the orientation of
the Sun's rotation axis and the difference between the port and
starboard ratios decreases as $|P|$ increases.

\subsection{Varying of the size of the solar image} 
\begin{figure}
  \centering
  \includegraphics[width=0.35\textwidth, clip]{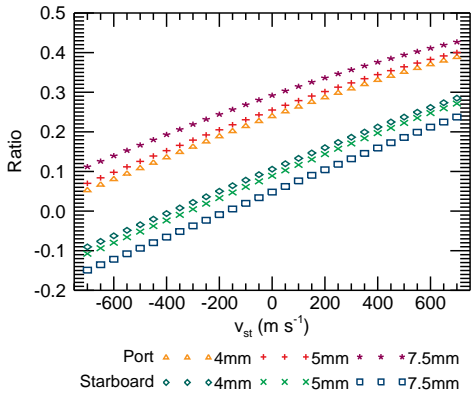}\\
  \caption{A comparison between the modelled ratio determined for
  different image radii. The modelled ratio is plotted as a
  function of station velocity. }\label{figure[wmv size comparison]}
\end{figure}

Although the
radius of the solar image observed in BiSON instruments is not
precisely known it is estimated to be of the order of $4-5\,\rm mm$.
Furthermore the observed radius of the Sun varies by $\sim3$ per
cent throughout the course of a year. Therefore the modelled ratio
at different values of $v_{\textrm{\scriptsize{st}}}$ was determined
when the size of the image was $4\,\rm mm$, $5\,\rm mm$ and
$7.5\,\rm mm$, which is the largest possible radius that the image
can have, assuming that the vapour cell observes the whole disc. We
have assumed the optical depth at the centre of the vapour cell,
$\tau_0$, is unity. Figure \ref{figure[wmv size comparison]} shows
that reducing the size of the observed image decreases the
difference between the port and starboard ratios. Therefore it is
necessary to know the size of the image in the vapour cell if
rigourous models of the observed weighted mean velocity are to be
made. Even the yearly variation of $3$ per cent in the size of the
solar image is non-negligible and so must be included when modelling
variation in the ratio with time.

\subsection{Varying the optical depth of the vapour}\label{subsection[tau change]} We believe
that the optical depth, $\tau_0$, in the vapour cell is of the order
of unity. However, the precise optical depth at the centre of the
cell is very difficult to determine observationally. Additionally,
any small variation in the temperature of the vapour in the cell
will alter its optical depth. We therefore investigated the effect
of varying the optical depth at the centre of the cell by altering
the value of the extinction coefficient, $K$. Calculations were made
when the optical depth at the centre of the cell, $\tau_0$, was 0.8,
1.0 and 1.2. The results are plotted in Figure \ref{figure[wmv
optical depth]}. The radius of the solar image in the vapour cell
was $5\,\rm mm$. Changing the optical depth of the vapour has a
similar effect on the calculated ratio as altering the size of the
image. In fact, varying $\tau_0$ by a value of 0.2 is equivalent to
altering the radius of the image by $1\,\rm mm$. Hence it is also
important to know what the optical depth at the centre of the vapour
cell is when modelling the ratio observed by unresolved Doppler
velocity observations.

\begin{figure}
  \centering
  \includegraphics[width=0.35\textwidth, clip]{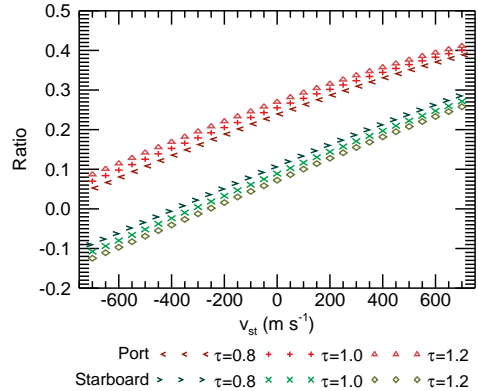}\\
  \caption{A comparison between the modelled ratio
  calculated when the optical depth, $\tau_0$, of the vapour is altered. The modelled ratio is plotted as a
  function of station velocity.}\label{figure[wmv optical depth]}
\end{figure}

\subsection{Determining the optical depth in the vapour cell and the
radius of the observed image} \label{subsection[find tau r]} 
\begin{figure}
  \centering
  \includegraphics[width=0.35\textwidth, clip]{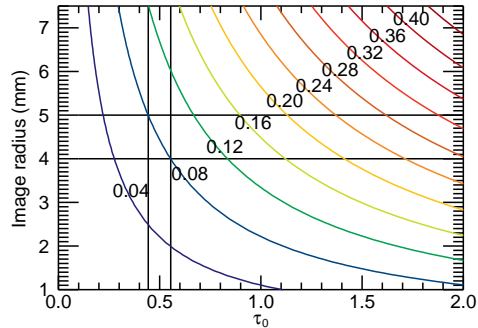}\\
  \caption{A contour map showing the variation in the difference between the port and
  starboard ratios with image radius and optical depth, $\tau_0$.}\label{figure[ratio difference]}
\end{figure}

Figure \ref{figure[ratio difference]}
is a contour map that shows the manner in which the separation of
the port and starboard ratios varies with changes to both the
optical depth at the centre of the cell and the image radius when
$v_{\textrm{\scriptsize{st}}}=0\,\rm m\,s^{-1}$. Although the
magnitude of the offset between the port and starboard ratios is
dependent upon which instrument is considered the difference between
the ratios is generally of the order of 0.08 when
$v_{\textrm{\scriptsize{st}}}=0\,\rm m\,s^{-1}$. Since the radius of
the solar image is believed to be in the range of $4-5\,\rm mm$,
Figure \ref{figure[ratio difference]} implies that the optical depth, $\tau_0$,
is in the range $0.55-0.44$ (as indicated by the black lines in
Figure \ref{figure[ratio difference]}).

We now use the observed daily ratios to determine a velocity offset
in the BiSON data and we use the derived information on the vapour
cell's optical depth to establish whether or not our model can
accurately replicate the observations.

\section{Velocity offset observed in BiSON data}\label{section[observed velocity offset]}

Equation \ref{equation[ratio2]} takes into account the varying
values of $v_{\textrm{\scriptsize{orb}}}$ and
$v_{\textrm{\scriptsize{spin}}}$ and so each value of
$R(v_{\textrm{\scriptsize{st}}})$ should correspond to the same
station velocity from day to day. We have determined the observed
velocity intercept that corresponds to a constant value of
$R(v_{\textrm{\scriptsize{st}}})$ for the port and starboard
detectors separately. The ratio, $R(v_{\textrm{\scriptsize{st}}})$,
was calculated using the daily best fit coefficients, $a_i$. We have
called the determined intercept the velocity offset.

The minimum observed ratio in any one day is often greater than
zero. Therefore, to determine the velocity that corresponds to a
ratio value of $R(v_{\textrm{\scriptsize{st}}})=0$ we would have to
extrapolate. However, the minimum observed ratio is often less than
$R(v_{\textrm{\scriptsize{st}}})=0.15$ and so the corresponding
velocity offset can be found by interpolation. Since the errors
associated with interpolation are less than the errors associated
with extrapolation we have chosen to find the velocity offset when
$R(v_{\textrm{\scriptsize{st}}})=0.15$.

For any one day the accuracy of the fit, which determines the
coefficients, $a_i$, is improved if more data points are observed.
Hence the velocity offset was only determined for days when more
than 500 ratios were observed since this constitutes approximately
half a day of data.

\begin{figure}
  \centering
  \subfigure{\includegraphics[width=0.35\textwidth,
  clip]{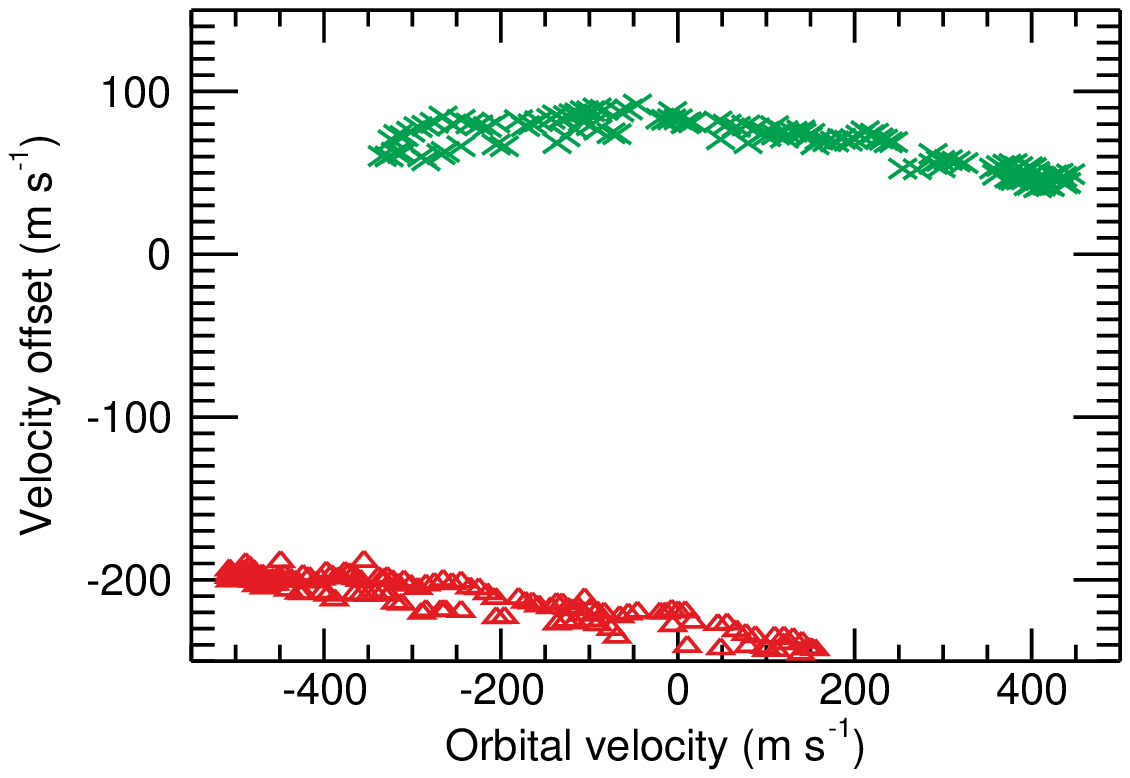}}\\
  \subfigure{\includegraphics[width=0.35\textwidth, clip]{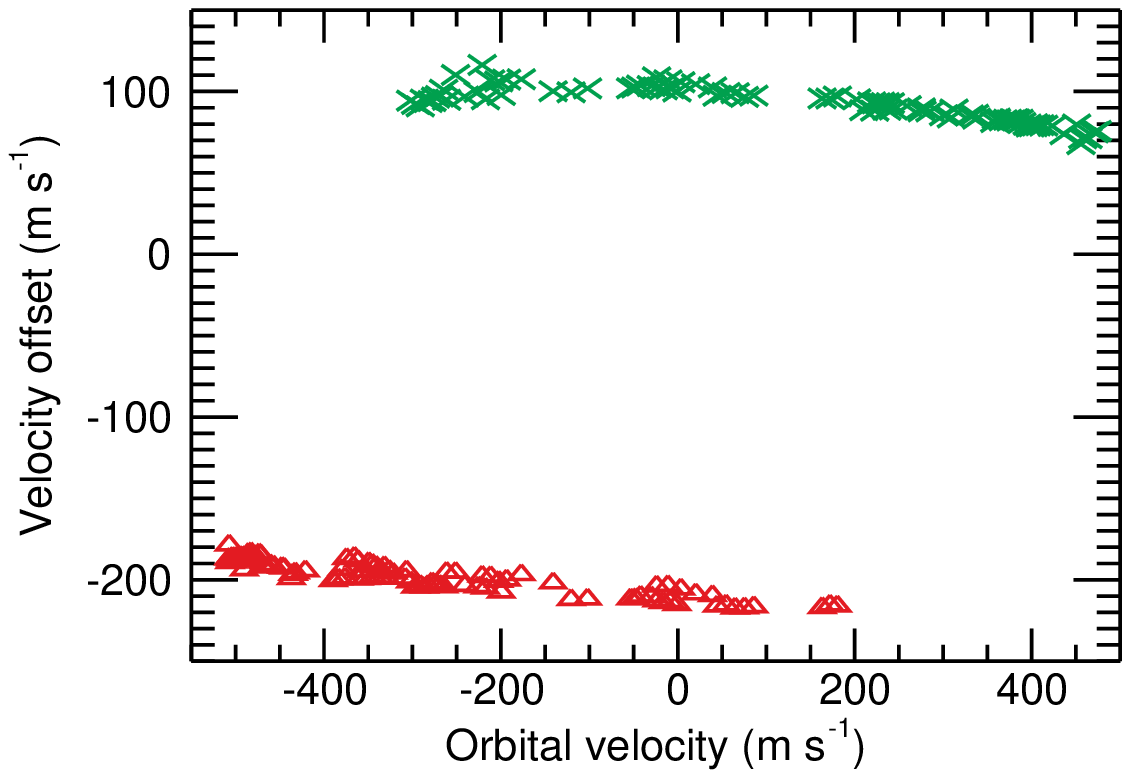}}\\
  \caption{Top panel: The velocity offset observed in Las Campanas data in 2006 as
  a function of $v_{\textrm{\scriptsize{orb}}}$. Bottom panel: The velocity offset observed in Narrabri data in 2006 as
  a function of $v_{\textrm{\scriptsize{orb}}}$. In each panel the red triangles represent
  the results from the port detector, while the green crosses show
  the velocity offset observed by the starboard detector.}\label{figure[obs vel offset]}
\end{figure}

Figure \ref{figure[obs vel offset]} shows the velocity offset
observed by two BiSON instruments as a function of the orbital
velocity, $v_{\textrm{\scriptsize{orb}}}$. The analysis was
performed for different instruments (and, although not shown here,
different years) to ensure the observed effects were not instrument
(or year) specific. The results are very similar for both sites and
the offset is observed to vary in a similar manner from year to
year. Figure \ref{figure[obs vel offset]} shows that the value of
the velocity offset not only varies with
$v_{\textrm{\scriptsize{orb}}}$, but the magnitude of the offset and
the shape of the variation are dependent on which detector is
considered. It should be noted that the offset observed in the mean
of the port and starboard data shows less variation than when the
two detectors are considered individually.

The process for determining the velocity offset in the BiSON data
was repeated using model data that takes into account the bias
across the disc due to the solar rotation, limb darkening and the
position of the detector. When creating the model data the optical
depth at the centre of the vapour cell was taken to be 0.55. The
radius of the image was varied throughout the year in a sinusoidal
manner and had a maximum radius, in January, of $4\,\rm mm$. The
orientation of the model Sun's rotation axis was varied sinusoidally
between $\pm26.3^\circ$ throughout the simulated year.

\begin{figure}
  \centering
  \includegraphics[width=0.35\textwidth, clip]{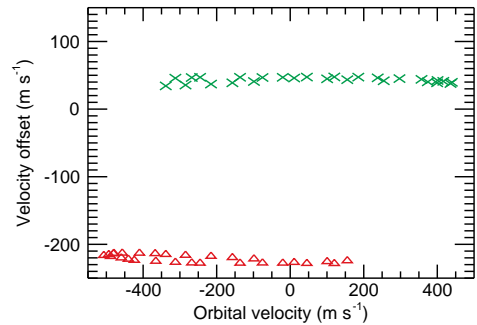}\\
  \caption{The modelled velocity offset as a function of
  $v_{\textrm{\scriptsize{orb}}}$. The red diamonds represent the
  results from the port detector, while the green crosses show the
  velocity offset observed by the starboard detector.}\label{figure[model offset]}
\end{figure}

Figure \ref{figure[model offset]} shows the modelled velocity
offset. Comparison between Figure \ref{figure[obs vel offset]} and
Figure \ref{figure[model offset]} indicates that the model is able
to replicate the approximate magnitudes of the offsets, the
separation between the port and starboard offsets and the general
shape of the offsets. However, the variation in the modelled port
offsets is approximatively a factor of 2 smaller than the variation
in the observed port offsets and the same is true for the starboard
offsets. The variation that is visible in the modelled offsets is
introduced primarily because the orientation of the Sun's rotation
axis varies with time. It is possible that a systematic variation in
other parameters such as the temperature, and consequently the
optical depth, of the vapour in the cell could also contribute to
the observed variation in the offsets. Furthermore, any variations
in the orientation of the image of the Sun in the vapour cell
because of instrumental effects will alter the observed offsets. The
presence of these offsets in the BiSON data is important for the
long term stability of the data, which ultimately affects the
quality of the data.

\section{Discussion}\label{section[discussion]}
We have shown that the observed Doppler imaging across the solar
disc is different for the red and blue wing observations and the
port and starboard observations. This means that the observations
are biased towards different portions of the solar Fraunhofer line.
From the centre of a solar absorption line to its wings the height
of formation in the solar atmosphere varies by hundreds of
kilometres. Therefore the port and starboard detectors measure
velocities at different depths in the solar atmosphere and so the
realization of the solar noise observed by each detector will be
only partially correlated. Furthermore, the observed amplitudes of
solar oscillations are dependent on the depth in the atmosphere at
which the observations are made and consequently the amplitudes of
the modes observed by each instrument will be different. The
variation in observation depth could impact on models of mode
excitation as the excitation rate is dependent on the mode mass,
which, to correctly match the observations must be determined at the
height at which the observations are made \citep{Baudin2005}.

Another consequence of the observed weighting across the solar disc
is the effect on the visibility of the modes. The visibility of a
mode varies depending on the sensitivity of the observing
instrument's response over the solar disc to a mode's eigenfunction.
\citet{Christensen1989} considered the manner in which the
visibilities of modes in unresolved observations are affected by the
solar rotation and limb darkening. Since the position of the
detector has a significant influence on the weighting across the
solar disc this too should be considered when determining the
visibility of modes. Furthermore, as the observed bias is dependent
on $v_{\textrm{\scriptsize{st}}}$, which varies with time, the
visibility of the modes will also be time dependent.

The observed Doppler imaging is sensitive to many different input
parameters. However, we have shown that the instruments have been
well calibrated to be sensitive to small changes in the
line-of-sight velocity. We have been able to use the separation of
the port and starboard ratios to estimate the Doppler imaging in the
cell. The difference between the port and starboard ratios in the
modelled and observed data are in agreement when the size of the
solar image in the vapour cell is $4\,\rm mm$ and the optical depth
at the centre of the vapour cell, $\tau_0$ is 0.55. The true values
of the size of the image and the optical depth in the vapour cell
are not known precisely. As the observed ratio is sensitive to the
optical depth (see Figures \ref{figure[wmv optical depth]} and
\ref{figure[ratio difference]}) this work highlights the need to
determine the value of the optical depth in the vapour cell and work
within the BiSON group is ongoing to this effect.

The sensitivity of the Doppler imaging to input parameters such as
the position angle of the Sun's rotation axis means that a velocity
offset is introduced into the BiSON data. Moderate agreement has
been achieved between the observed and modelled velocity offsets.
However, we believe there is a further effect that is related to a
variation in the tilt of the instrument that could explain the
difference between the observations and the model. The presence of a
velocity offset introduces low-frequency noise to the BiSON data. We
have shown that this offset can potentially be modelled and can,
therefore, in principle be removed from the data, thereby reducing
the observed level of noise. The introduced noise has a 1/frequency
dependence and so it is particularly important at very low
frequencies. Therefore, removing this noise source make it easier to
observe low-frequency p modes and g modes in Sun-as-a-star data.

\section*{Acknowledgements}

We are grateful to R. Simoniello for allowing us to use Themis data.
This paper utilizes data collected by the Birmingham
Solar-Oscillations Network (BiSON). We thank the members of the
BiSON team, both past and present, for their technical and analysis
support. We also thank P. Whitelock and P. Fourie at SAAO, the
Carnegie Institution of Washington, the Australia Telescope National
Facility (CSIRO), E.J. Rhodes (Mt. Wilson, Californa) and members
(past and present) of the IAC, Tenerife. BiSON is funded by the
Science and Technology Facilities Council (STFC). The authors also
acknowledge the financial support of STFC.

\bibliographystyle{mn2e}
\bibliography{weighting_paper_v3}

\end{document}